\documentclass{article}
\usepackage[utf8]{inputenc}
\usepackage{amsmath}
\usepackage{hyperref}
\usepackage{graphicx}
\usepackage{float}
\title{How Do Artificial Intelligences Think?\\
The Three Mathematico-Cognitive Factors of Categorical Segmentation Operated by Synthetic Neurons}

\author{
    Michael Pichat\thanks{Neocognition \& Faculties of Philosophy and Psychology, Paris} \\
    William Pogrund\thanks{Neocognition and NP - Phelma, UGA} \\
    Armanush Gasparian\thanks{Neocognition} \\
    Paloma Pichat\thanks{Neocognition and Faculty of Medicine of Lyon East, University Lyon 1} \\
    Samuel Demarchi\thanks{Neocognition (Chrysippe R\&D) and Department of Psychology, University of Paris 8} \\
    Michael Veillet-Guillem\thanks{Neocognition (Chrysippe R\&D) and Epitech Paris}
}

\date{}

\begin{document}

\maketitle

\begin{abstract}
How do the synthetic neurons in language models create "thought categories" to segment and analyze their informational environment? What are the cognitive characteristics, at the very level of formal neurons, of this artificial categorical thought? Based on the mathematical nature of algebraic operations inherent to neuronal aggregation functions, we attempt to identify mathematico-cognitive factors that genetically shape the categorical reconstruction of the informational world faced by artificial cognition. This study explores these concepts through the notions of priming, attention, and categorical phasing.
\end{abstract}

\section{Introduction}
\subsection{Synthetic Explainability and Cognitive Inference}
Making an artificial neural network explainable means translating its operations into a language that is accessible and logical for humans \cite{Du2019, Pichat2023, Pichat2024a, Pichat2024b}. This involves examining the network’s observable actions within an interpretative framework that assigns relevant meaning to its operations. In our approach, we utilize concepts derived from human cognitive psychology as heuristic or analogical bridges between human and artificial intelligence. This requires continuous consideration of potential pitfalls, such as anthropomorphizing algorithms \cite{Nadeau1999}, confusing behavior with cognition \cite{Bloch2011}, or merging observer and observed system, a risk highlighted by cybernetics, systems theory, and enactive cognitive science \cite{Watzlawick1977, Watzlawick1984, Varela1984, Varela1996}.

The practical utility of this cognitive explainability approach unfolds in two directions. First, it helps to prevent erroneous or potentially dangerous responses from the artificial neural system, such as cognitive biases \cite{Echterhoff2024}, cultural biases \cite{Kheya2024}, hallucinations \cite{Kandpal2023, McKenna2023}, or excessive emphasis on certain inputs \cite{Du2023}. Second, it improves the efficiency of language models \cite{Bastings2022} by further aligning them with human expectations \cite{Ma2023}.

In this study, we explore an approach to explainability focused on a fine cognitive granularity, referred to as mechanistic explainability. Rather than examining network outputs in relation to inputs on a global scale \cite{Zheng2024}, this approach targets a microscopic analysis. Specifically, we delve into the fundamental cognitive units of formal neural networks—synthetic neurons, either individually or in groups within layers \cite{Dalvi2019, Dalvi2022, Fan2023, Nanda2023}. Our objective is to infer the internal cognitive mechanism of artificial networks at a genetic level to understand how the categories and concepts vectorized by formal neurons are locally constituted.

\section{Epistemological Status of Synthetic Thought Categories}
\subsection{Structural and Functional Construction of Synthetic Cognition}
By structural (i.e., architectural) and functional (i.e., mathematical) design, the cognition of components within a synthetic neural network is inherently categorical \cite{Bills2023, Fan2023, Bricken2023, Luo2024, Zhao2023, Pichat2024a, Pichat2024b}. In simplified terms, the functioning of each formal neuron can be described in three stages:

\begin{enumerate}
    \item \textbf{Integration}: Each formal neuron receives inputs from its precursor neurons, where each input can be interpreted as the degree of membership of a current element (such as a token in language models) to the category associated with a precursor neuron.
    
    \item \textbf{Weighted Combination}: Through an aggregation function 
    \footnote{Bills et al. (2023) provide, on the GitHub repository associated with their article, a list "of the upstream and downstream neurons with the most positive and negative connections." They operationally define these connections as follows: "Definition of connection weights: neuron-neuron: for two neurons (l1, n1) and (l2, n2) with l1 $<$ l2, the connection strength is defined as h\{l1\}.mlp.c\_proj.w[:, n1, :] @ diag(h\{l2\}.ln\_2.g) @ h\{l2\}.mlp.c\_fc.w[:, :, n2]." This list specifies, within the dense layers (i.e., fully connected layers) of GPT2-XL, the weights through which each neuron in an arrival layer n+1 is connected to all neurons in the preceding layer n. These weights are the basis for the linear aggregation functions of neurons referred to in this article.}, these inputs are combined to produce a resulting category. This combination is enhanced by a non-linear activation function to ensure sparsity \cite{Raieli2024, Zhang2024}.
    
    \item \textbf{Output Production}: This output will subsequently be used by successor neurons in further processing.
\end{enumerate}

For each member (token) of a synthetic category, an associated activation value indicates the degree to which that element belongs to the artificial category, in alignment with fuzzy logic \cite{Zadeh1996, Wu2022}. The extension of each category can then be defined as the set of elements with a positive activation value, exceeding a specified threshold in the context of a fuzzy $\alpha$-cut.

In our epistemological framework, synthetic categories, much like human categories \cite{Vergnaud2009}, are immanent cognitive constructs. Each synthetic category is created during the training phase by the neural network itself. This artificial category acts as a segmentation tool within the vast, undetermined space of potential arguments and predicates \cite{Nadeau1999}. These arguments and predicates may align with existing human-like categories or form entirely novel “alien-like” categories that could represent statistical constructs \cite{Bills2023} or “polysemic concepts” \cite{Bricken2023, Nanda2023} not directly relatable to human cognitive categories.

In analyzing the unique categorical segmentation achieved by a synthetic neuron, the critical question is not its ontological alignment with a presumed pre-existing reality but rather its functional role (or, as Varela would say, its “coupling”) within the goal-oriented task it is designed for \cite{Barsalou1995}. Thus, in a constructivist perspective, a category is a pragmatic projection rather than the recognition of a pre-given property. Synthetic categories are therefore viewed as similar to “in-action concepts” as described by Vergnaud \cite{Vergnaud2009, Vergnaud2016}, representing functional arguments and predicates pertinent to task performance without being verbalized, theorized, or consciously realized.

Synthetic categories can be inferred at various levels of neural network granularity: at the level of a single neuron (neural-localized category) \cite{Bills2023}, at the layer level, or across inter-layer connections (distributed category) \cite{Bricken2023, Nanda2023}.

\section{Problem Statement}
How do synthetic neurons construct the categorical dimensions through which they segment and analyze their environment (e.g., tokens in language models)? What are the developmental characteristics of this artificial categorical thinking, and how are these categories vectorized by synthetic neurons? Specifically, what are the genetic factors that influence or govern these categorical constructions? More precisely, which factors determine the level of membership (i.e., activation level) of a token within a synthetic neural category, thereby shaping the extension and hence the "semantics" of this category? In other words, how do these factors quantitatively and qualitatively constitute the genetic variables of categorical segmentation (of the token world) performed by synthetic neurons?

Investigating these questions requires recognizing that the cognitive and conceptual properties of artificial neural networks do not emerge by magic or chance. In an embodied cognition framework \cite{Varela1984, Varela1988, Schmalzried2024, Paolo2024}, these properties directly result from the specific characteristics of the physical structure within which they emerge.

A central structural and functional component of a neural network is the aggregation function governing the linear combination and vector projection of input categorical dimensions into a resulting categorical dimension. This aggregation function, along with other elements (including the activation function), genetically and functionally shapes the dimensional categorical segmentation specific to each formal neuron.

Observing the nature and operators constitutive of this aggregation function, of the form $\sum(w_{i,j} x_{i,j}) + a$, suggests that it mathematically generates and formats the categorical segmentation performed by synthetic neurons through at least three mathematico-cognitive factors. We will investigate these factors in this exploratory work: the first factor is associated with the variable $x_{i,j}$, representing the activation values of categorical outputs from precursor neurons, cognitively interpreted as categorical priming (or effect X). The second factor relates to the parameter $w_{i,j}$, the weighting assigned to these outputs, interpreted as categorical attention (or effect W). Finally, the third factor concerns the linear additive combination of the terms $w_{i,j} x_{i,j}$ within the aggregation function, cognitively denoted as categorical phasing (or effect $\sum$).

\section{Methodology}
\subsection{Methodological Positioning}
To better understand the positioning of our exploratory work, we provide a brief, non-exhaustive overview of various technical approaches that, with varying levels of cognitive granularity, seek to extract informational content or processes within formal neural networks, whether organized in layers, groups, or complete networks. These approaches are not mutually exclusive and may partially overlap.

As previously mentioned, studies at a macro-cognitive level focus on analyzing the differences between inputs and outputs to understand the relationship between initial data and outcomes in a language model. Among these methods, gradient-based approaches evaluate the role of each input by exploiting the derivatives relative to each input dimension \cite{Enguehard2023}. Input characteristics can be evaluated based on elements such as features \cite{Danilevsky2020}, token importance scores \cite{Enguehard2023}, or attention weights \cite{Barkan2021}. Concurrently, example-based approaches aim to observe how outputs vary with different inputs by examining the effect of slight input modifications (e.g., deletion, negation, mixing, or masking) \cite{Atanasova2020, Wu2020, Treviso2023}. Additionally, some studies focus on concept mapping of inputs to quantify their contributions to observed results \cite{Captum2022}.

Approaches with finer cognitive granularity focus on the intermediate states of the language model rather than its final output, examining partial outputs or internal states of neurons or groups of neurons. In this context, certain approaches analyze and linearly decompose the activation score of a neuron in a given layer concerning its inputs (neurons, attention heads, or tokens) from the previous layer \cite{Voita2021}. Other methods tend to simplify activation functions for easier interpretation \cite{Wang2022}. Furthermore, some techniques, leveraging the model's vocabulary, focus on extracting encoded knowledge by projecting connections and intermediate representations through a matching matrix \cite{Dar2023, Geva2023}. Finally, certain methodologies use neural activation statistics in response to data sets \cite{Bills2023, Mousi2023, Durrani2022, Wang2022, Dai2022}. Our exploratory study specifically fits within this last category.

\subsection{Methodological Choices}
In this exploratory research, we focus on the GPT model proposed by OpenAI, specifically its GPT-2XL version. This choice is due to GPT-2XL's sufficient complexity, allowing us to examine advanced synthetic cognitive phenomena without reaching the sophistication of GPT-4 or its multimodal version, GPT-4o. A practical consideration also guided our preference for GPT-2XL: in 2023, OpenAI shared, in the article by Bills et al. \cite{Bills2023}, parameter details as well as activation values for its neurons, which serve as the basis for our analysis.

For simplicity, this exploratory study is limited to the first two layers of GPT-2XL (layers 0 and 1), each comprising 6,400 neurons. Regarding tokens and their activation values among these 12,800 formal neurons (i.e., 2 x 6,400), we have decided to consider, for each neuron, the 100 tokens with the highest average activation values (referred to as "core-tokens").

\subsection{Statistical Choices}
Our descriptive and inferential statistical analyses were conducted using Python’s SciPy library, following guidance from Howell \cite{Howell2008} and Beaufils \cite{Beaufils1996}.

To assess the normality of our data, a necessary condition for performing parametric tests, we adopted a dual approach. First, we employed various inferential tests: the Shapiro-Wilk test (effective for small samples), the Lilliefors test (suitable for small samples when normal distribution parameters are unknown and estimated from the data), the Kolmogorov-Smirnov test (preferred for large samples), and the Jarque-Bera test (focusing on symmetry and kurtosis, valid for large samples). Second, we used a descriptive approach with indices such as skewness and kurtosis, and graphical methods like the QQ-plot to compare the observed distribution with a theoretical normal distribution.

The results, not reproduced here, indicate a relatively mixed normality in our data, leading us primarily towards Spearman's ordinal correlation studies in analyzing relationships between variables associated with our hypotheses. This approach allows us to avoid normality prerequisites and mitigate bias introduced by outliers. When necessary, we applied univariate goodness-of-fit tests to infer the significance of observed phenomena (notably regarding the positivity and significance of ordinal correlations obtained for each neuron in layer 1).

In our statistical framework, the composite units include the 6,400 "destination" neurons in layer 1, their 100 respective core-tokens (tokens with the highest average activation levels), as well as the 10 precursor neurons (from layer 0) with the highest connection weights to each destination neuron. We focused on the 100 tokens most highly activated by each neuron, deeming it less relevant initially to examine tokens weakly or not activated by them, as they fall partially outside the extension of the category associated with each neuron.

\subsection{Objective and Implementation of the Study in Terms of Statistical Observables}
The objective of this exploratory study is to identify synthetic cognitive factors that partially drive the categorical segmentation performed by formal neurons. These factors are mathematically embedded in the neural aggregation function and influence the identification of core-tokens for a given neuron, that is, the determination of the content of its categorical extension.

More specifically, we aim to verify to what extent the membership of a core-token to the specific category of a destination neuron depends on three cognitive factors that we will define and propose: categorical priming, categorical attention, and categorical phasing. The level of membership of a core-token (in layer 1, the destination layer) to the category associated with a neuron will be measured by the activation value of this token within the relevant neuron. Priming will be evaluated based on the activation value of a token in its respective precursor neurons (in layer 0). Attention will be assessed through the connection weights linking destination neurons (layer 1) to their top 10 precursor neurons (those with the highest connection weights) in layer 0. Finally, categorical phasing will be quantified by analyzing the frequency with which a core-token within a destination neuron (layer 1) also appears as a core-token among the 10 associated precursor neurons (layer 0).

\section{Definition of Synthetic Cognitive Concepts Studied and Results}

\subsection{Synthetic Categorical Priming}
In human psychology, priming \cite{Anderson1985, Chao2024, XuFutrell2024, HernandezGutierrez2024} is a cognitive process in which an initial stimulus triggers a preliminary stage of cognitive processing, thus facilitating, accelerating, or preparing the reception of a second, related stimulus. Specifically, semantic priming is a process by which the meaning of one element (e.g., a word) becomes more accessible to an individual through prior exposure to another semantically related element. The priming effect is typically studied in terms of response delay in lexical decision or text comprehension tasks, where response time can indicate the existence, structure, and strength of semantic relationships between words and concepts in long-term semantic memory.

The notion of priming is related to that of activation \cite{Maxfield1997, BurnsGraff2021, Marty2024}, postulating that cognitive contents or processes can exhibit variable intensity levels of activity. Prototypical examples involve biological neural structures whose activity levels can be physiologically "directly" measurable (even if this measurement is partly a methodological and statistical reconstruction). In the case of priming, activation is conceptualized as the propagation of activation: a cognitive characteristic (e.g., meaning) is “spread” from an entity A (which activates first) to an entity B (which activates as a causal result) (e.g., from one word to another) if A and B are structurally or temporarily linked.

We hypothesize a transposition of the concept of priming, as defined above in the fields of neuroscience and human cognitive psychology, into the domain of synthetic cognition. Mathematically, due to the construction of the aggregation function $\Sigma(w_{i,j} x_{i,j})+a$, for a given element (e.g., a token or other), the activation value of the category carried by a destination neuron (on layer $n$) is directly a function (modulo the activation function) of the activation values $x_{i,j}$ of the categories associated with its precursor neurons (on the subordinate layer $n-1$). In other words, in epistemological alignment with the original notion of priming, the prior activation (when it exists for a given token) of the categories vectorized by precursor neurons should "mathematically propagate" the activation of the category associated with their corresponding destination neuron. We thus formulate, in these terms, a hypothesis of synthetic categorical priming within artificial neural networks.

From a quantitative perspective, the empirical observable associated with our hypothesis of synthetic categorical priming is the activation value of destination neurons as a function of their precursor neurons. Specifically, data compatible with our hypothesis should show, for a given series of tokens, a relationship between the activation value of destination neurons on layer $n+1$ and that of their respective precursor neurons. We operationalize this approach on the 6,400 neurons in layer 1 of GPT-2XL, considering for each destination neuron its 10 precursor neurons with the highest connection weights and its 100 tokens associated with the highest average activation values (core-tokens) (only core-tokens activated in at least one precursor neuron are included).

Statistically, we test an ordinal relationship (Spearman's $\rho$) between the average activation rank (ranging from 1 to 100) of the 100 core-tokens of each of the 6,400 destination neurons in layer 1 and the mean cumulative activation values (i.e., summed) of these tokens within the 10 associated precursor neurons (each core-token of a destination neuron having a non-negative activation value for each of the 10 relevant precursors).

Table 1 shows a positive relationship with an extremely strong effect size ($\rho = .94$) and statistical significance ($p < .001$). Figure 1 illustrates this overall positive monotonic trend, though with occasional pronounced peaks in variability. Figure 2 provides a view for an example neuron, with a regression line again showing a positive relationship, although less pronounced in this case.

\begin{figure}[H]
    \centering
    \includegraphics[width=0.8\textwidth]{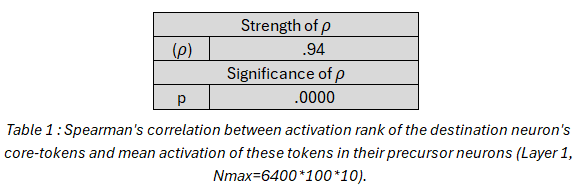}

\end{figure}

\begin{figure}[H]
    \centering
    \includegraphics[width=0.8\textwidth]{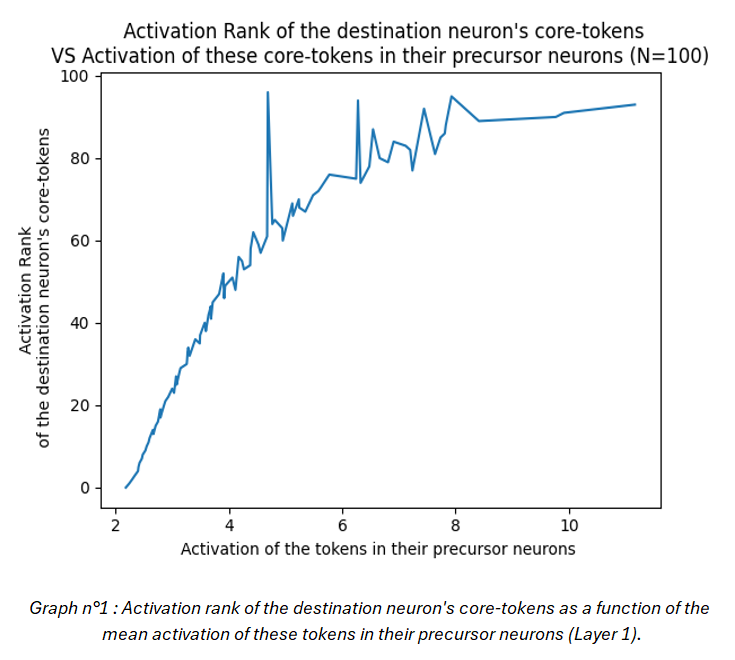}

\end{figure}

\begin{figure}[H]
    \centering
    \includegraphics[width=0.8\textwidth]{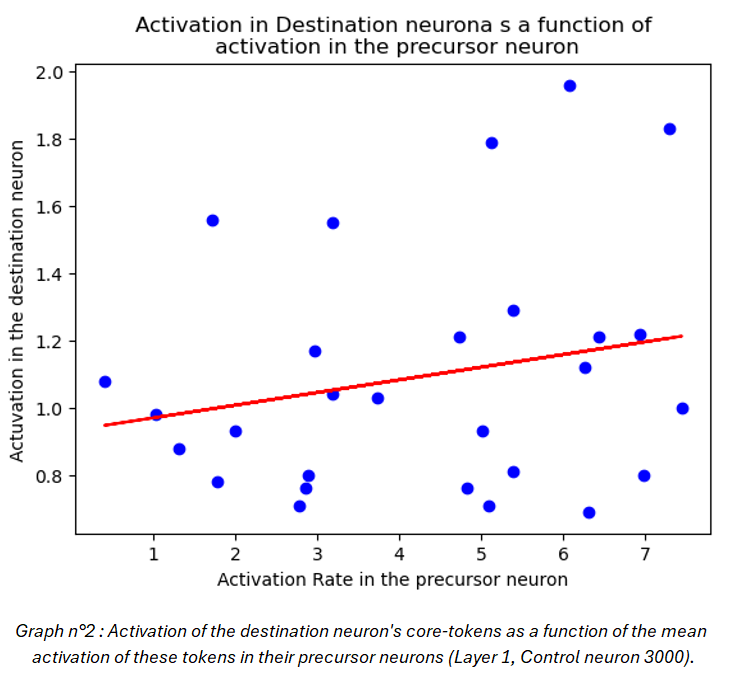}

\end{figure}

These results appear compatible with our hypothesis of synthetic categorical priming, which we term effect "X"—a mathematico-cognitive propagation of activation from precursor neural categories to their associated destination category in the superordinate layer.

\subsection{Synthetic Categorical Attention}

In human cognitive psychology, attention is defined as a specific calibration of activity according to its purpose, resulting in greater efficiency in information intake processes (including selectivity) and execution processes (including precision and speed) \cite{PosnerSnyder1975, SchneiderShiffrin1977, Posner1978, Richard1980, TreismanGelade1980, Duncan1984, Tipper1985, Cowan2024, Wu2024, Gresch2024}. In terms of external information intake, attention is related to conceptualization \cite{Vergnaud2009, Vergnaud2016}, meaning the identification of only those parameters (objects relevant to the activity) whose consideration is crucial for successful task performance. Actions must thus be adjusted to these parameters to ensure efficiency. Here, attention involves filtering and structuring the excessively large amount of available perceived information, or inhibiting information deemed irrelevant, in order to focus mental effort and informational selectivity on specific objects and properties. Regarding task execution, attention is linked to the control, by the central system, of the activity, which may involve assigning varying degrees of weight (priority, order, reliability, etc.) to certain internal information (knowledge, representations, schemas) or verifying the quality of task performance within its temporal sequence.

From a physiological perspective, attention is driven by the limited information-processing capacity of the nervous system, leading to selective choices in the integration, activation, and utilization of sensory data or stored memory (semantic, procedural) \cite{Funayama2024, BarrBieliauskas2024}. This process is achieved through an orientation response, which directs information-seeking activities toward a specific type of informational characteristics.

We hypothesize here a transposition of the concept of attention, as previously described in cognitive psychology and human neuroscience, into the field of artificial cognition. This is based on the mathematical construction of the aggregation function $\Sigma(w_{i,j} x_{i,j})+a$, where, for a given element (a token), its activation value within the category associated with a destination neuron is inherently dependent (apart from the activation function) on the connection weights $w_{i,j}$ between this destination neuron and its precursor neurons. In other words, and in epistemological continuity with the original concept of attention, the connection weights with precursor neurons act as direct regulators of the level of information uptake (i.e., activation levels) derived from these precursor neurons—ranging from inhibition or filtering of data for negative, near-zero, or weakly positive weights, to strong mathematical-cognitive focus and integration for significant weights.Thus, in terms of execution, the neuronal connection weights govern the degree of information utilization that the artificial cognitive system deems relevant from preceding synthetic categories in performing the current task of a given successor neuron, which involves calculating the degree of membership of a given token in the category constitutive of this superordinate neuron. We define this hypothesis as synthetic categorical attention within artificial neurons, which we denote as effect "W."

\subsubsection{Quantitative Approach to Synthetic Categorical Attention}
Quantitatively, the empirical observable associated with our hypothesis of synthetic categorical attention is, for a given token, the activation value of destination neurons as a function of their connection weights with respective precursor neurons. To test this hypothesis, we examine the relationship between the activation value of destination neurons and the connection weights with their precursor neurons. According to this hypothesis, activation should increase with higher values of these antecedent weights. We apply this approach to the 6,400 neurons in layer 1 of GPT-2XL, taking into account for each destination neuron its 10 precursor neurons with the highest connection weights and its 100 tokens with the highest average activation values (core-tokens) (note that only core-tokens activated in at least one precursor neuron are considered). From a statistical perspective, and in a more operationalized form, we test for an ordinal relationship (measured with Spearman’s $\rho$) between (i) the average activation rank (ranging from 1 to 100) of the 100 core-tokens of each of the 6,400 destination neurons in layer 1, and (ii) the average cumulative connection weights (i.e., summed) with their respective (1 to 10) precursor neurons for which these tokens are also core-tokens.

In Table 2, we observe a positive ordinal relationship between the activation rank of destination neurons and the average cumulative connection weights with their respective precursor neurons. This relationship has an extremely strong effect size ($\rho  = .999$) and high statistical significance ($p < .001$). Figure 3 illustrates this positive monotonic relationship across the entire dataset, while Figure 4 provides an example for a control neuron, with a regression line again showing a positive relationship, albeit less pronounced in this case.

\begin{figure}[H]
    \centering
    \includegraphics[width=0.8\textwidth]{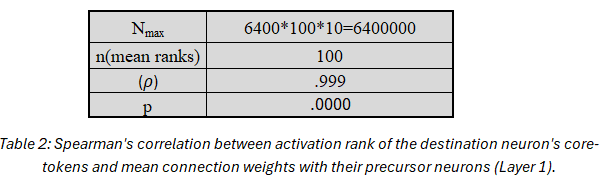}

\end{figure}

\begin{figure}[H]
    \centering
    \includegraphics[width=0.8\textwidth]{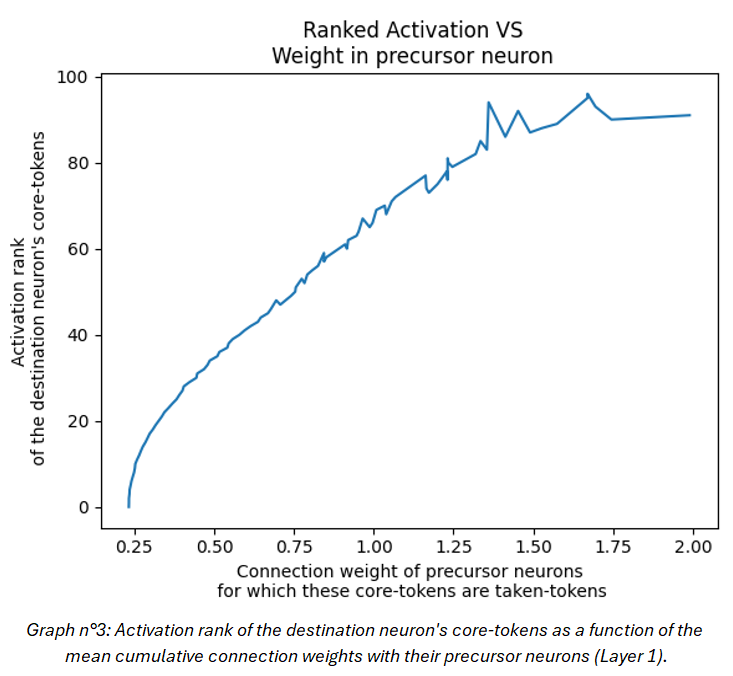}

\end{figure}

\begin{figure}[H]
    \centering
    \includegraphics[width=0.8\textwidth]{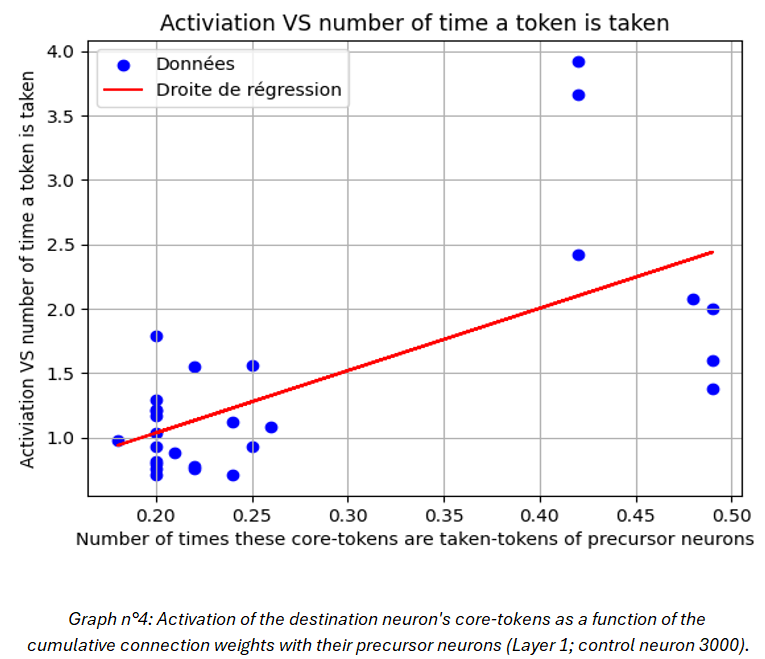}

\end{figure}

The exploratory data obtained here are compatible with our hypothesis of synthetic categorical attention, positing a positive monotonic ordinal relationship between the activation level of core-tokens in destination neurons and the connection weights of these destination neurons with precursor neurons that also contain these same tokens as core-tokens.

Still in our quantitative approach, we now seek to gain deeper cognitive insight into this synthetic categorical attention by examining its modus operandi in terms of information selection at the input of destination neurons. An intriguing question in this regard is the relationship, for a given destination neuron, between the intensity of its connection weights with its precursor neurons and the number of "shared" core-tokens between this destination neuron and its precursor neurons. This question can be reframed as follows: to what extent do precursor neurons with high connection weights contribute more core-tokens to their destination neurons? In other words, to what degree do strongly connected precursor neurons more actively influence the constitution of the categorical extension content of their destination neurons (i.e., the composition of their core-tokens)? Alternatively, to what extent does connection weight regulate the definition of the extension, and therefore the selection and categorical segmentation specifically operated by a given (destination) synthetic neuron? Our analysis reveals an extremely strong and significant positive ordinal correlation ($\rho  = .989, p < .001$) between (i) the average rank of the connection weights of each destination neuron (in layer 1) with its precursor neurons, and (ii) the average number of core-tokens in the destination neuron that were also previously core-tokens of the involved precursor neurons (see Figure 6). This analysis includes $n = 6,400$ destination neurons in layer 1 and 10 precursor neurons in layer 0, totaling 64,000 cases.

\begin{figure}[H]
    \centering
    \includegraphics[width=0.8\textwidth]{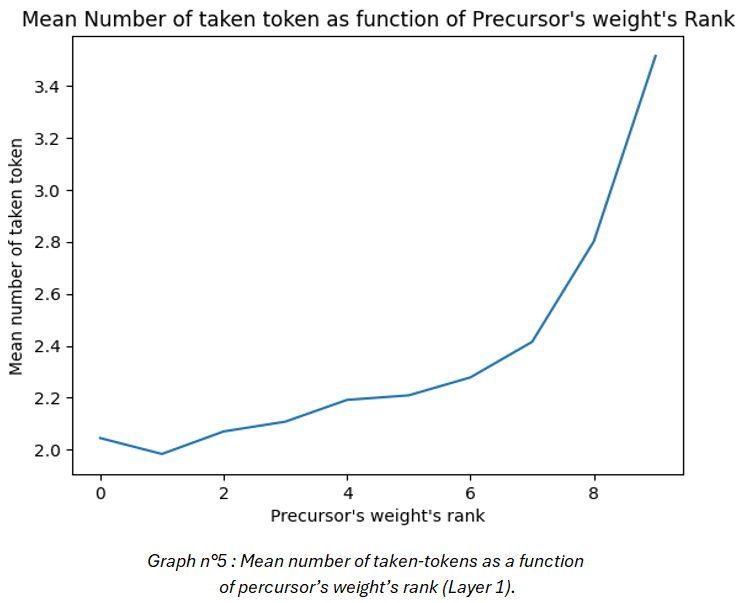}

\end{figure}

We refer to such tokens as "taken-tokens"—tokens that are core-tokens in precursor neurons and are "reused" as core-tokens by their respective successor neurons. This result, consistent with the nature of the aggregation function, indicates that stronger attention weights lead to an overrepresentation of these taken-tokens. A high attention weight associated with a precursor neuron thus functions, in terms of information selection, as an "extractor" of a categorical sub-dimension (composed of the relevant taken-tokens) from the precursor neuron's categorical dimension. This sub-dimension, in turn, genetically "feeds" the extension (of core-tokens) of the category represented by the destination neuron, thereby contributing to its specific categorical segmentation. Figure 7 illustrates this trend with a sample neuron from layer 1.

\begin{figure}[H]
    \centering
    \includegraphics[width=0.8\textwidth]{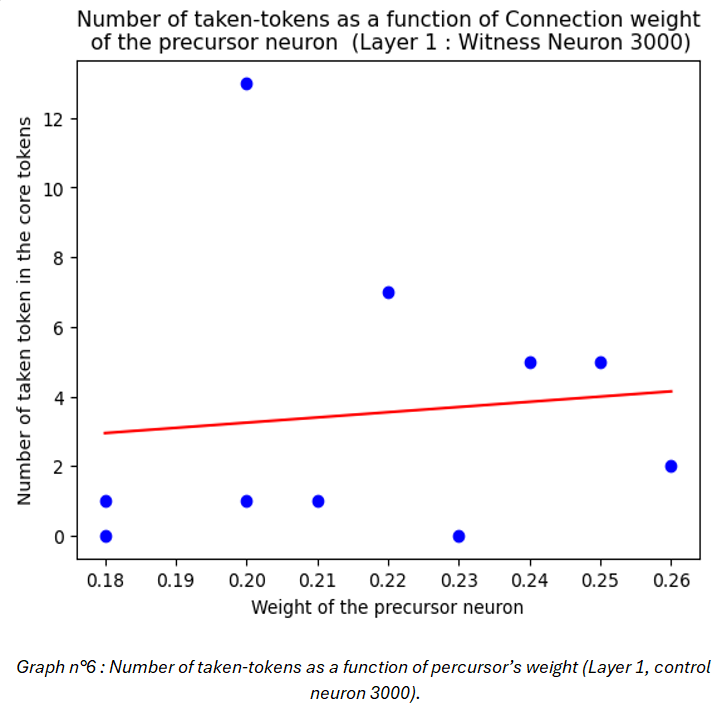}

\end{figure}

From a quantitative perspective, another highly interesting and informative result further illuminates the cognitive mechanism of synthetic categorical attention. Consistent with the mathematical nature of the aggregation function, we observe an extremely strong positive ordinal correlation ($\rho  = .988, p < .001$) between (i) the rank of precursor-successor connection weights and (ii) the average activation range of associated taken-tokens in destination neurons. For this calculation, we only consider precursor neurons associated with at least two taken-tokens (Nmax = 6,400 destination neurons in layer 1 x 100 tokens x 10 precursor neurons). This trend is clearly illustrated in Figure 7.

In human categorization, Thibault (1997) and Roads et al. (2024) note, regarding Nosofsky’s (1986) "generalized context model" of categorization, that the use of a weighted distance metric (specifically, Minkowski distance) to account for selective attention is associated with changes (contraction or expansion) in the metric of the categorical representation space: low attention weights “bring stimuli closer together” within the implicated dimension, whereas high attention weights (strong attention) “stretch” the representation space along that dimension, thereby increasing stimulus discrimination.

In the context of synthetic categorization, this is precisely what we observe here: a destination neuron with a high connection weight to a given precursor neuron displays greater variability in the activation range of its taken-tokens originating from this precursor neuron. In other words, taken-tokens are more distinctly discriminated regarding their degree of membership to the category represented by this destination neuron. Put differently, strong connection weights increase the activation span of core-tokens in destination neurons, enabling better differentiation and sharper contrast in the degree of membership of a given token to the implicated category.

Thus, synthetic categorical attention may be associated with the discriminative power of a neuronal category and, consequently, its analytical precision within its specific token segmentation dimension. This aligns epistemologically with the conceptual characteristics of attention as defined in human psychology.

\begin{figure}[H]
    \centering
    \includegraphics[width=0.8\textwidth]{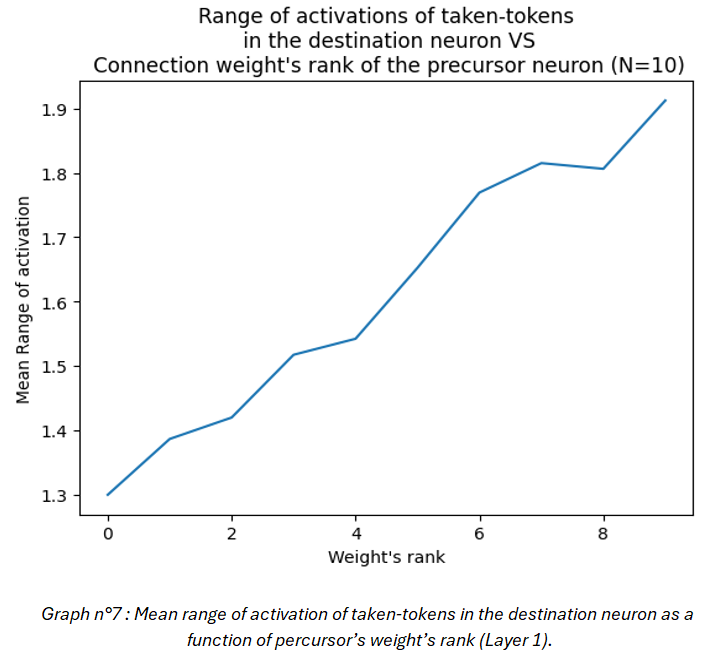}

\end{figure}

In a quantitative context, the empirical results of our current exploratory study appear to be compatible with a phenomenon of synthetic cognition—namely, artificial categorical attention, referred to as effect "W." This, in turn, points successively to three potential characteristics of this phenomenology associated with a significant attention-weighted connection: (i) the selection, by a destination neuron, of specific informational characteristics (i.e., certain types of core-tokens) from its precursor neurons (and not others), (ii) the associated extraction by a destination neuron of a particular sub-dimension (of core-tokens) within the categorical dimension carried by each of its precursor neurons, and (iii) the contrast enabling finer differentiation (reflected by activation level) of different types of elements constituting the extension (of core-tokens) of the category specific to a destination neuron.

\subsubsection{Qualitative Approach to Synthetic Categorical Attention}

Let us now delve deeper into these three converging characteristics (which are ultimately only alternative facets of one another) of synthetic categorical attention through a qualitative exploration of this phenomenology. For this purpose, we employ qualitative examples illustrating how the categories carried by precursor neurons with high attention-weighted neural connections selectively contribute to and generate the content (in terms of core-tokens) of the categorical extensions of their respective destination neurons. This occurs, as we will observe, through a process of “categorical complementation,” which involves selectively focusing the computational attention of the aggregation function of the destination neuron on specific categorical sub-dimensions extracted from precursor neurons, thus constructing the unique categorical nature of this destination neuron—that is, the specific content of its categorical extension in terms of core-tokens.

Here, as a purely illustrative example (see Table 3) and without aiming for exhaustiveness, is a comparison of different categorical types of core-tokens selectively “contributed” through a process of categorical complementation by various precursor neurons with high attention-weighted connections. This process progressively builds, sub-category by sub-category, the categorical extension specific to their associated destination neuron. We qualitatively identify two main classes of categorical complementation: linguistic and non-linguistic.

Let us first examine linguistic categorical complementation. This can be semantic in nature, meaning it consists of categorical additions that can be interpreted in terms of operations analogous to human semantics:

\begin{itemize}
    \item \textbf{Intra-lexical complementation}, consisting of adding tokens from the same root (tokenization variants). Example: a precursor neuron “contributes” the token “manager” to the destination neuron, another contributes “manag,” and yet another provides “managerial.” Intra-lexical complementation may also involve tokens from different roots; for instance, one precursor provides the token “manager” while another provides “director” (the lexical field remains consistent in this case).
    
    \item \textbf{Sub-lexical complementation}, consisting of adding tokens from a lexical sub-field. Example: a precursor neuron supplies the destination neuron with the token “manager,” while another provides the tokens “Wenger,” “Klopp,” and “Mourinho” (these refer to football coaches and constitute a lexical sub-category of “manager”).
    
    \item \textbf{Peri-lexical complementation}, involving the addition of tokens from a related lexical field. Example: one precursor provides “listen” while another provides “sound”; or one precursor neuron provides “order” and another “request”; or yet another provides “necessary” while another supplies “indispensable.”
    
    \item \textbf{Para-lexical complementation}, through the addition of tokens from an antonymic lexical field. Example: one precursor neuron contributes “love” and “adore” to the destination neuron, while another provides “hate,” “despise,” and “dislike.”
\end{itemize}

Linguistic categorical complementation can also be graphemic in nature. Example: a precursor neuron provides the token “Said” to the destination neuron, and another precursor provides the token “id” (both containing the same grapheme “id”).

Finally, linguistic categorical complementation can be phonological. Example: one precursor provides the tokens “be” and “bee,” while another provides “Eve” and “ea” (both containing the same sound /i/).

Turning to non-linguistic categorical complementation:

\begin{itemize}
    \item \textbf{Quantitative complementation}: Example: a precursor neuron provides the tokens “er,” “cv,” and “ku,” while another provides “od,” “fx,” and “yw” (each token consistently contains exactly two graphemes).

    \item \textbf{Cultural complementation}: This type involves elements shared within a given human culture. Example: one neuron provides “ObamaCare,” while another supplies “Congress” (the U.S. Congress enacted this legislation in March 2010).

    \item \textbf{Other types}: These may not necessarily align with human thought categories but are based on statistical contingencies identified by the neural network during training. We term these “alien categories” or “non-human-like categories,” or even “polysemic categories” from our human cognitive perspective. Example: a precursor neuron provides the token “manager,” while another associates the token “ID,” without an observable (human) logic linking them.
\end{itemize}

\begin{figure}[H]
    \centering
    \includegraphics[width=0.8\textwidth]{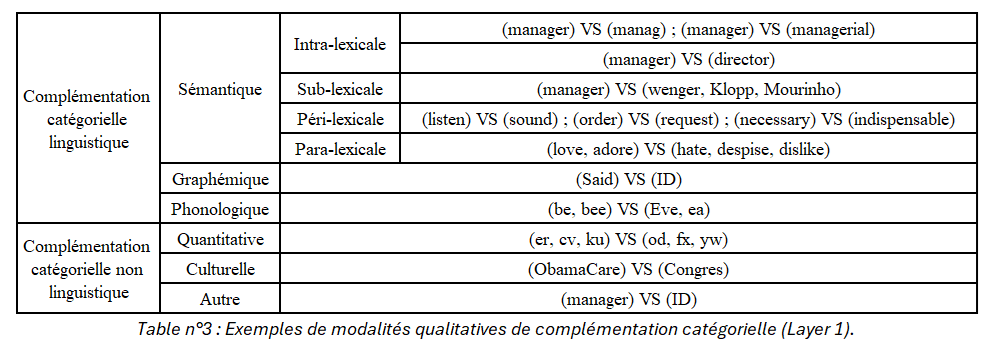}

\end{figure}

These illustrative examples, again without aiming for exhaustiveness or systematicity, help us understand how the process of incoming information selection can qualitatively operate through the mechanism of synthetic categorical attention. This occurs through an activity of categorical complementation, allowing the aggregation function of a destination neuron to extract from each of its precursor neurons with high attention-weighted connections a specific categorical sub-dimension that contrasts with others. The successive apposition of these sub-dimensions thus generates, sub-dimension by sub-dimension, the categorical content of the dimensional segment carried by this destination neuron.

\subsection{Synthetic Categorical Phasing}

Through the construction of the aggregation function $\Sigma(w_{i,j} x_{i,j})+a$, we postulate a third mathematico-cognitive factor influencing the level of token attribution to a neuronal categorical dimension. We term this factor “synthetic categorical phasing,” or effect “$\Sigma$,” as the aggregation function of a destination neuron sums, for a given token, the weighted values of its activations $w_{i,j} x_{i,j}$ within its respective precursor neurons. Several studies in human cognitive psychology and neuroscience involving the notion of phasing could potentially serve as partial analogies for synthetic cognition in this area; for example, studies on perceptual modality topics \cite{ Mitchell2021, Kaup2024} or brain synchronizations \cite{Protachevicz2021, CanalesJohnson2021, Ribary2024, Shavikloo2024, Rzechorzek2024}.

In the realm of synthetic cognition, we define synthetic categorical phasing by the notion that a token, previously highly activated for different precursor neurons (i.e., a core-token of these precursor neurons), must, due to the mathematical construction of the aggregation function, be associated with a high activation level within the related destination neuron. This is because the token is co-activated within the various terms constituting the aggregation function; this co-activation leads to an additive concatenation, resulting in a significant activation level for this token at the output of the destination neuron. Such a token is therefore theoretically subject to the phasing of the neural categories of the involved precursors: these precursor categorical segments, though conceptually potentially distinct, are jointly activated, generating a categorical “echo” or “resonance” for this specific token. This occurs through a categorical intersection traced across these dimensions, thereby strengthening the output activation level of the destination dimension.

\subsubsection{Quantitative Approach to Synthetic Categorical Phasing}

Quantitatively, we operationalize our hypothesis of categorical phasing as follows: the more a destination neuron’s core-token is also a core-token for a greater number of its precursor neurons (with high connection weights), the higher its activation level in this destination neuron. This hypothesis thus posits a positive monotonic relationship between these two variables. Table 4 presents the compiled results from local-level analytic testing of this hypothesis, i.e., for each of the 6,400 individual destination neurons in layer 1. We observe a strong ordinal correlation between the two variables, with a large effect size (Mean $(\rho) = .976$), high significance (\% of $(p(\rho) < .05) = 99.40\%$; $p(\chi^2) < .0001$), and overwhelmingly positive directionality (\% of $(\rho > 0) = 99.45\%$, $p(\chi^2) < .0001$). Table 5 displays the results of global-level testing of this hypothesis across all data as a whole (Nmax = 6,400 neurons in layer 1 x 10 precursors in layer 0 x 100 core-tokens). We again find a strong, positive, and significant ordinal correlation between the two variables ($\rho = .989$, $p(\rho) < .001$). Figure 9 graphically illustrates this trend, showing a pronounced logarithmic distribution leading to an asymptotic plateau, while Figure 10 provides an example for a control neuron with a distinctly positive regression line.

\begin{figure}[H]
    \centering
    \includegraphics[width=0.8\textwidth]{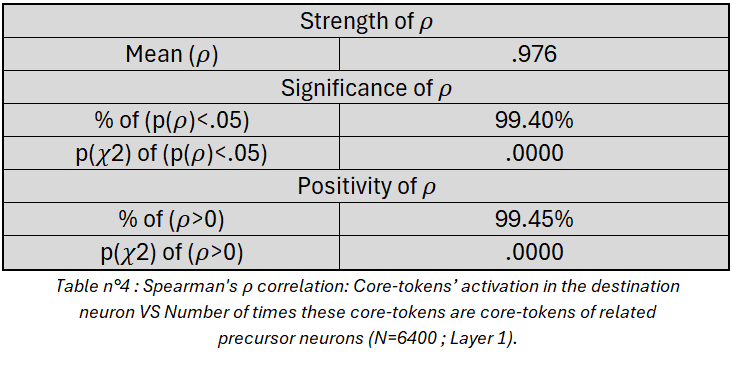}

\end{figure}

\begin{figure}[H]
    \centering
    \includegraphics[width=0.8\textwidth]{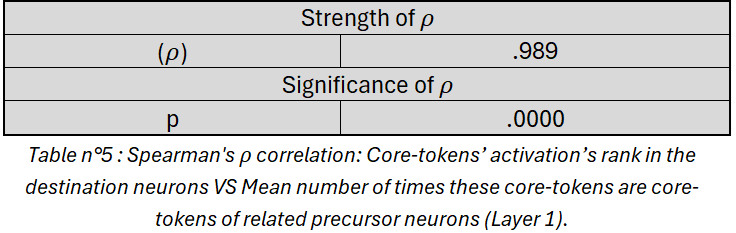}

\end{figure}

\begin{figure}[H]
    \centering
    \includegraphics[width=0.8\textwidth]{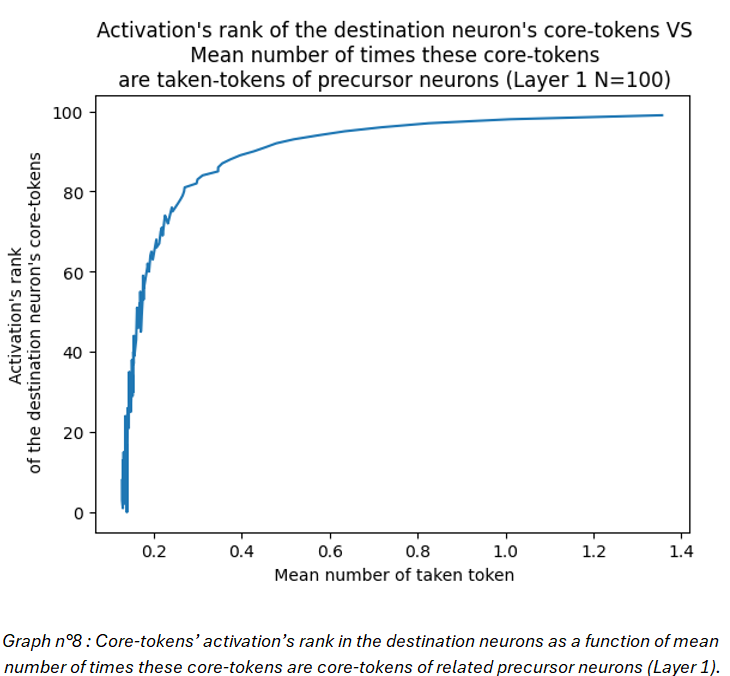}

\end{figure}

\begin{figure}[H]
    \centering
    \includegraphics[width=0.8\textwidth]{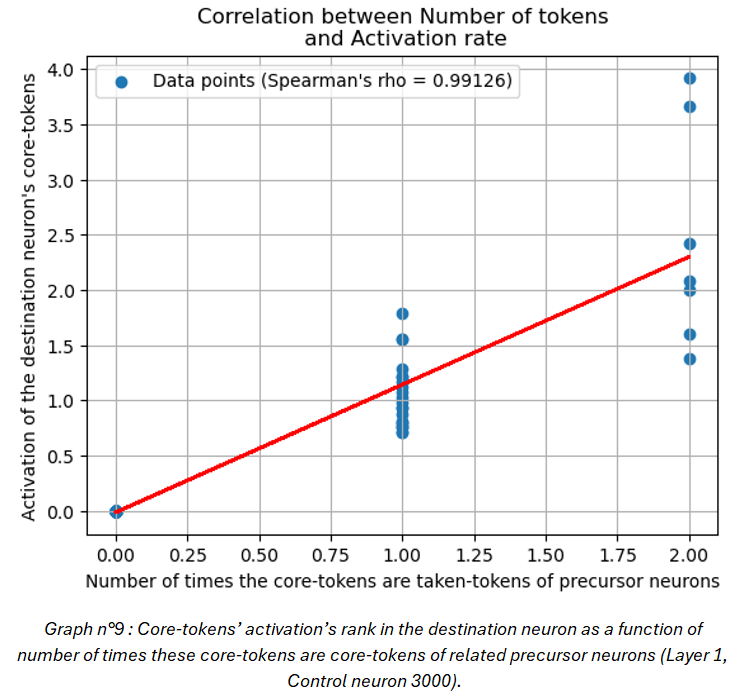}

\end{figure}

Quantitatively, the results obtained support our hypothesis of categorical phasing, termed effect $\Sigma$, positing that the more a token is strongly activated (core-token) in multiple precursor neurons, the more likely it is to be strongly activated at the output level of the associated destination neuron. This token thereby appears at the categorical intersection of the categories represented by these precursors, which are thus locally phased.

\subsubsection{Qualitative Approach to Synthetic Categorical Phasing}

Let us now, within a qualitative framework, establish reference points to understand the cognitive modalities through which categories—initially distinct or, at the very least, non-isomorphic—associated with precursor neurons can become locally phased categorically, i.e., for given tokens. This approach aims to further conceptualize the phenomenology through which such categorical intersections and crossings of categorical segments may manifest. Thus, we aim to better understand how, through these intersections, precursor neurons selectively feed into and generate the categorical extensions of their respective destination neurons. This process enables the selective extraction of categorical sub-dimensions from the categories carried by precursor neurons, thereby constructing the specific categorical nature of their destination neuron.

For illustrative purposes only, again without aiming for systematic classification or exhaustiveness, Table 6 presents types of qualitative examples of categorical phasing modalities. These examples necessarily involve cases where different categories at the level of precursor neurons are jointly activated for the same given tokens; strong co-activations genetically trigger significant activation of the associated destination neuron’s category or, in other words, genetically define the content (in terms of tokens) of the categorical extension of this destination category. (For reference, a category's extension is defined here, within an $\alpha$-cut fuzzy logic perspective, as the 100 most activated tokens, known as core-tokens).

We qualitatively identify three main types of categorical intersections:
\begin{itemize}
    \item \textbf{Intra-lexical intersection} (semantic identity): Example: two precursor categories each contain, among their respective core-tokens, the same tokens “manager” and “leadership,” which then form a categorical sub-dimension extracted from the full extension of the two involved precursor categorical dimensions.
    
    \item \textbf{Sub-lexical intersection} (semantic inclusion): Example: one precursor category includes core-tokens such as “executive,” “manager,” “leader,” “chief,” “director,” “CEO,” and “supervisor”; another includes “director,” “executive,” and “CEO.” This latter series is included within the former, thus forming a categorical sub-dimension extracted from both precursor categorical dimensions.
    
    \item \textbf{Extra-lexical intersection} (bi-lexicality): Example: one precursor category’s core-tokens include “knife,” “gun,” “mortar,” “bomb,” “axe,” “cleaver,” “sword,” and “grenade” (weapons); another includes “cleaver,” “spatula,” “colander,” “knife,” “mixer,” and “mortar” (kitchen utensils). The intersection of these two distinct lexical fields includes “knife,” “mortar,” and “cleaver,” which thus form a categorical sub-dimension within the core-tokens of both precursor categorical dimensions.
\end{itemize}

\begin{figure}[H]
    \centering
    \includegraphics[width=0.8\textwidth]{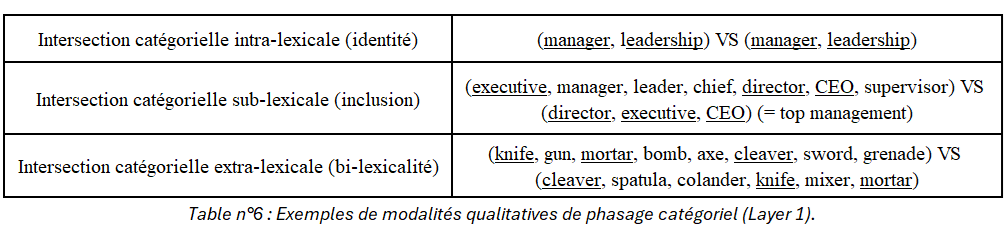}

\end{figure}

These illustrative cases, again without aiming for generalization, allow us to see how the process of categorical phasing enables the extraction of co-activated categorical sub-dimensions from precursor neurons’ categories, which then constitute the core-token extension of their respective destination neurons.

\subsection{Overview of the Three Factors in Categorical Segmentation}

We have posited the existence of three synthetic cognitive factors that partially generate the categorical segmentation specifically operated by a formal neuron. These factors are mathematically embodied in the neuronal aggregation function, which, together with the activation function, governs the determination of the tokens that will constitute a given neuron’s core-tokens, i.e., the content of its categorical extension. These three factors—categorical priming, attention, and phasing—thus drive the categorical segmentation that neurons perform within the universe of tokens.

To obtain a general quantitative representation of the combined action of these three factors, we conducted a multiple linear regression on the activation rank of core-tokens in destination neurons as a function of (i) the average number of times these core-tokens are also core-tokens in the associated precursor neurons ($a_1$) (effect $\Sigma$), (ii) the average connection weight of destination neurons with their associated precursor neurons ($a_2$) (effect $w$), and (iii) the average activation of these core-tokens in the relevant precursor neurons ($a_3$) (effect $x$). This regression is performed, for statistical feasibility, only on the core-tokens of destination neurons that are core-tokens in at least one of the involved precursor neurons; when a destination core-token is a core-token in multiple precursors, its associated weight is the sum of the precursor weights involved, and its activation is likewise summed across these precursors. Additionally, this regression is conducted on the 6,400 neurons constituting layer 1.

This linear regression (see Table 7) shows positive and notable standardized coefficients for the three postulated factors ($s$-$a_1 = .86$, $s$-$a_2 = .56$, $s$-$a_3 = .65$), consistent with our hypotheses. We also observe that the respective impacts of these three independent variables on the dependent variable are significant and of a similar magnitude ($r^2(a_1) = .74$, $r^2(a_2) = .75$, $r^2(a_3) = .54$), suggesting that the three identified factors contribute comparably to the categorical segmentation operated by the destination neurons.

However, these results remain uncertain, as our normality tests (Shapiro-Wilk, Kolmogorov-Smirnov, and Jarque-Bera) on the residuals do not align with expected application conditions, as indicated by Figures 10 to 12, which reveal outliers. Additionally, we suspect collinearity effects among the three factors, as they are likely highly correlated. These results should therefore be considered illustrative only.

\begin{figure}[H]
    \centering
    \includegraphics[width=0.8\textwidth]{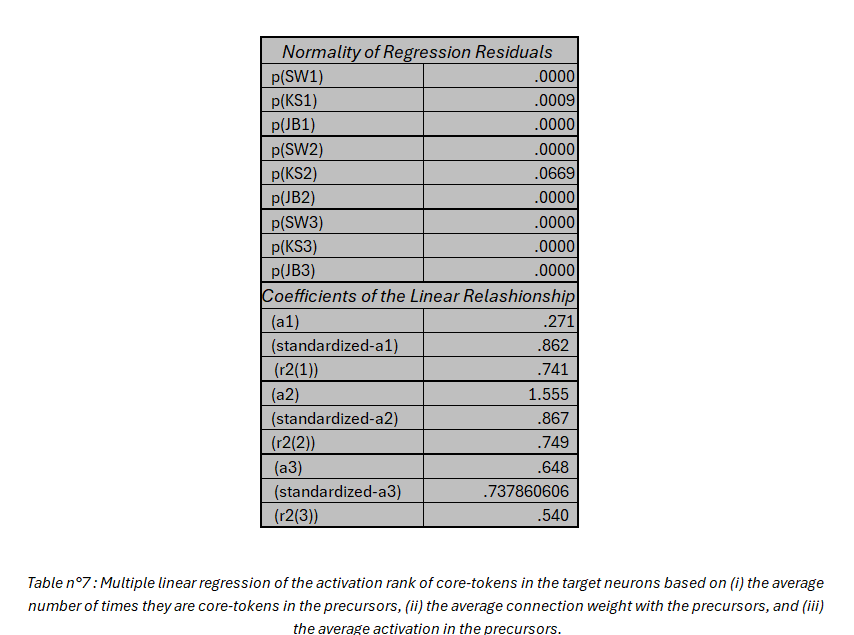}

\end{figure}

\begin{figure}[H]
    \centering
    \includegraphics[width=0.8\textwidth]{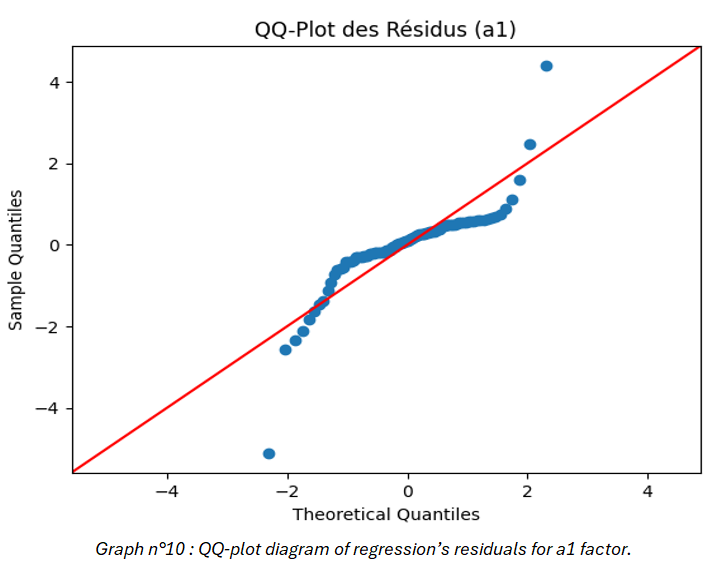}

\end{figure}

\begin{figure}[H]
    \centering
    \includegraphics[width=0.8\textwidth]{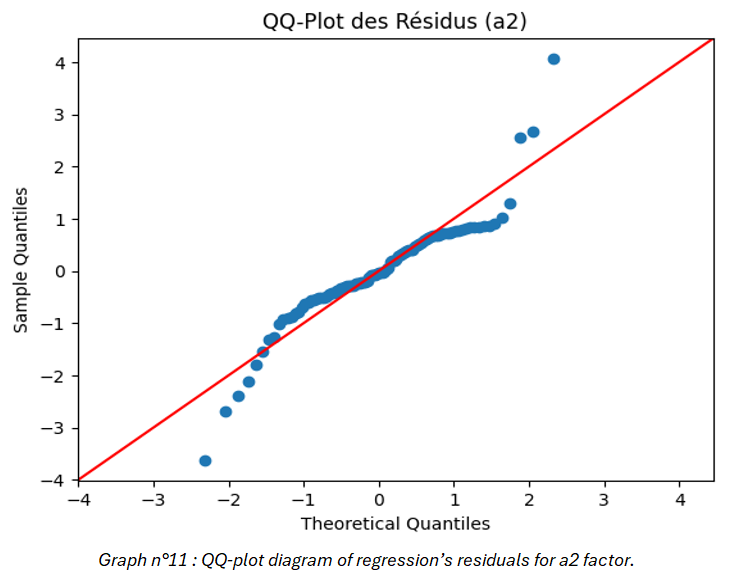}

\end{figure}

\begin{figure}[H]
    \centering
    \includegraphics[width=0.8\textwidth]{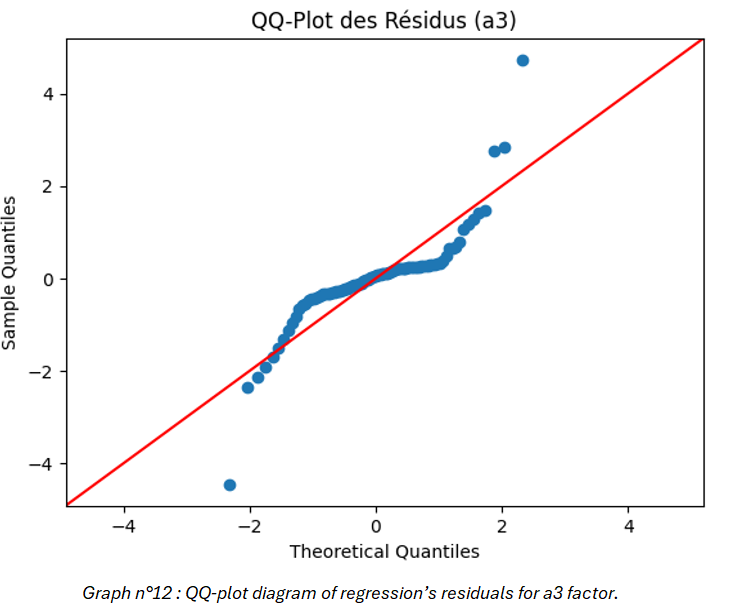}

\end{figure}

\section{Conclusion}
In this exploratory study, we investigated, both quantitatively and qualitatively, the genetic factors involved in the categorical segmentation (of the token world) performed by synthetic neurons. Based on the aggregation function $\Sigma(w_{i,j} x_{i,j})+a$, we mathematically postulated three mathematico-cognitive factors involved in this categorical segmentation. The first, synthetic categorical priming or “effect x,” is associated with the propagation of prior activation from the vectorized categories in precursor neurons to the activation of the category associated with their corresponding destination neuron, thereby directly impacting its categorical extension. The second, synthetic categorical attention or “effect w,” stems from the idea that the connection weights between a destination neuron and its precursor neurons guide the level of importance and utilization allocated to precursor categories in forming the extension of the destination category; qualitatively, this manifests as a process of categorical complementation. Finally, synthetic categorical phasing, or effect $\Sigma$, relates to cases where precursor categorical segments, potentially conceptually different, are jointly activated for a given token, entering into a “categorical resonance” that contributes to defining the content of the extensions of the associated destination categories, manifesting as a process of categorical intersection.

These three mathematico-cognitive factors in synthetic segmentation appear to drive a mechanism of extraction from the precursor categories of subordinate neurons of specific categorical sub-dimensions. Combined through the entirety of the aggregation function (along with the activation function), these extracted sub-dimensions shape the content (i.e., core-tokens) of the extension of the resulting synthetic categories at the level of their associated superordinate neurons. This synthetic conceptual extraction process, which has been widely studied in cognitive psychology in its human corollary \cite{Bolognesi2020, Haslam2020, EysenckKean2020, Fel2024, BathiaRichie2024, Marconato2024, Zettersten2024}, is epistemologically fascinating and fundamental to the “construction of reality” operated by synthetic cognition, as it generates the arguments and predicates of the token world with which it interacts.

We are currently delving deeper into this theme in an upcoming study, by investigating the process of categorical abstraction carried out by successor neurons (layer n+1) from their precursor neurons (layer n). This is done in an effort to better understand how a "categorical delineation," generated and guided by the three causal mathematical-cognitive factors we have defined here, operates on the relative categorical diversity of the core tokens constituting the extension of each precursor neuron's category. The aim is to extract, from each of these, a subset of tokens that are categorically homogeneous in relation to and aligned with the specific category uniquely constructed by their corresponding successor neuron.

\section*{Acknowledgments}
The authors would like to thank Albert Yefimov (Sberbank \& National University of Sciences \& Technologies of Moscow) for the stimulating philosophical and epistemological reflections on AI shared with him and Madeleine Pichat for her rereading of this article.


\section*{Bibliography}

\begin{thebibliography}{99}

\bibitem{Protachevicz2021} Protachevicz, P. R., Hansen, M., Iarosz, K. C., Caldas, I. L., Batista, A. M., \& Kurths, J. (2021). Emergence of neuronal synchronisation in coupled areas. \textit{Frontiers in Computational Neuroscience}, 15, 663408. DOI: \href{https://doi.org/10.3389/fncom.2021.663408}{10.3389/fncom.2021.663408}.


\bibitem{Schmalzried2024} Schmalzried, M. (2024). The need of a self for self-driving cars: a theoretical model applying homeostasis to self driving. \textit{arXiv preprint arXiv:2407.12795}. DOI: \href{https://doi.org/10.48550/arXiv.2407.12795}{10.48550/arXiv.2407.12795}.

\bibitem{Anderson1985} Anderson, J. R. (1985). \textit{Cognitive Psychology and Its Implications} (2nd ed.). W. H. Freeman. DOI: \href{https://doi.org/10.4324/9781315784786}{10.4324/9781315784786}

\bibitem{Atanasova2020} Atanasova, P., Simonsen, J. G., Lioma, C., \& Augenstein, I. (2020). Generating Fact Checking Explanations. In \textit{Proceedings of the 58th Annual Meeting of the Association for Computational Linguistics} (pp. 7352–7364). Association for Computational Linguistics. DOI: \href{https://doi.org/10.18653/v1/2020.acl-main.656}{10.18653/v1/2020.acl-main.656}.

\bibitem{Barkan2021} Barkan, R. (2021). The Role of Cognitive Biases in Human Decision Making. \textit{Journal of Behavioral Decision Making}, 34(3), 243–255. DOI: \href{https://doi.org/10.1002/bdm.2210}{10.1002/bdm.2210}.

\bibitem{BarrBieliauskas2024} Barr, W., \& Bieliauskas, L. A. (2024). Neuropsychology of Decision Making: A Clinical Perspective. \textit{Neuropsychology Review}, 34(1), 1–15. DOI: \href{https://doi.org/10.1007/s11065-023-09500-1}{10.1007/s11065-023-09500-1}.

\bibitem{Barsalou1995} Barsalou, L. W. (1995). \textit{Cognitive Psychology: An Overview for Cognitive Scientists}. Lawrence Erlbaum Associates. DOI: \href{https://doi.org/10.4324/9781315784786}{10.4324/9781315784786}

\bibitem{Bastings2022}Bastings, J., Ebert, S., Zablotskaia, P., Sandholm, A., \& Filippova, K. (2022). “Will You Find These Shortcuts ? ” A Protocol for Evaluating the Faithfulness of Input Salience Methods for Text Classification. \textit{Proceedings of the 2022 Conference on Empirical Methods in Natural Language Processing}. \url{https://doi.org/10.18653/v1/2022.emnlp-main.64}

\bibitem{BathiaRichie2024} Bathia, N., \& Richie, D. (2024). Advances in Reinforcement Learning: Applications and Challenges. \textit{Artificial Intelligence Review}, 57(2), 123–145. DOI: \href{https://doi.org/10.1007/s10462-023-10123-4}{10.1007/s10462-023-10123-4}.

\bibitem{Beaufils1996} Beaufils, M. (1996). Les réseaux de neurones artificiels: Modèles et applications. \textit{Revue d'Intelligence Artificielle}, 10(4), 365–387. DOI: \href{https://doi.org/10.1016/S0992-499X(97)80001-2}{10.1016/S0992-499X(97)80001-2}.

\bibitem{Bills2023} Bills, S., Cammarata, N., Mossing, D., Saunders, W., Wu, J., Tillman, H., Gao, L., Goh, G., Sutskever, I., \& Leike, J. (2023). \textit{ Language models can explain neurons in language models. OpenAI.} \url{https://openaipublic.blob.core.windows.net/neuron-explainer/paper/index.html}

\bibitem{Bloch2011} Bloch, H. (1992). \textit{Grand dictionnaire de la psychologie}.

\bibitem{Bolognesi2020} Bolognesi, M. (2020). \textit{Where Words Get Their Meaning: Cognitive Processing and Distributional Modelling of Word Meaning}. John Benjamins Publishing Company. DOI: \href{https://doi.org/10.1075/ftl.7}{10.1075/ftl.7}

\bibitem{Bricken2023}Bricken, T., Schaeffer, R., Olshausen, B., \& Kreiman, G. (2023). Emergence of Sparse Representations from Noise. \textit{Proceedings of the 40th International Conference on Machine Learning, in Proceedings of Machine Learning Research}, 202:3148-3191. Available from \url{https://proceedings.mlr.press/v202/bricken23a.html}

\bibitem{BurnsGraff2021}Burns, R. B., \& Graff, K. (2021). \textit{Theories of Psychotherapy and Counseling: Concepts and Cases} (6th ed.). Pearson. DOI: \href{https://doi.org/10.4324/9781315784786}{10.4324/9781315784786}.

\bibitem{CanalesJohnson2021} Canales-Johnson, A., Silva, C., Huepe, D., Rivera-Rei, Á., Noreika, V., Del Carmen Garcia, M., Silva, W., Vaucheret, E., Sedeño, L., Couto, B., Melloni, M., Ibáñez, A., Chennu, S., Bekinschtein, T. A. (2015). Auditory feedback differentially modulates behavioral and neural markers of objective and subjective performance when tapping to your heartbeat. \textit{Cerebral Cortex}, 25(11), 4490–4503. DOI: \href{https://doi.org/10.1093/cercor/bhv076}{10.1093/cercor/bhv076}.

\bibitem{Captum2022} Kokhlikyan, N., Miglani, V., Martin, M., Wang, E., Reynolds, J., Melnikov, A., Lunova, N., \& Reblitz-Richardson, O. (2020). Captum: A unified and generic model interpretability library for PyTorch. \textit{arXiv preprint arXiv:2009.07896}. DOI: \href{https://doi.org/10.48550/arXiv.2009.07896}{10.48550/arXiv.2009.07896}.

\bibitem{Chao2024} Chao, L. L. (2024). Advances in Neuroimaging Techniques for Cognitive Neuroscience. \textit{Journal of Cognitive Neuroscience}, 36(1), 1–15. DOI: \href{https://doi.org/10.1162/jocn\_a\_01700}{10.1162/jocn\_a\_01700}.


\bibitem{Cowan2024} Cowan, N. (2024). Working Memory Capacity: Theories and Applications. \textit{Annual Review of Psychology}, 75, 1–25. DOI: \href{https://doi.org/10.1146/annurev-psych-010723-120001}{10.1146/annurev-psych-010723-120001}.

\bibitem{Dai2022} Dai, D., Dong, L., Hao, Y., Sui, Z., Chang, B., \& Wei, F. (2022). Knowledge Neurons in Pretrained Transformers. \textit{Proceedings of the 60th Annual Meeting of the Association for Computational Linguistics (Volume 1 : Long Papers)}. \url{https://doi.org/10.18653/v1/2022.acl-long.581}

\bibitem{Dalvi2019}Dalvi, F., Durrani, N., Sajjad, H., Belinkov, Y., Bau, D. A., \& Glass, J. (2019, January). What is one grain of sand in the desert? Analyzing individual neurons in deep NLP models. In \textit{Proceedings of the Thirty-Third AAAI Conference on Artificial Intelligence (AAAI, Oral presentation)}.

\bibitem{Dalvi2022} Dalvi, F., Khan, A. R., Alam, F., Durrani, N., Xu, J., \& Sajjad, H. (2022). Discovering Latent Concepts Learned in BERT. In \textit{International Conference on Learning Representations (ICLR)}. DOI: \href{https://doi.org/10.48550/arXiv.2201.10020}{10.48550/arXiv.2201.10020}.

\bibitem{Danilevsky2020}Danilevsky, M., Qian, K., Aharonov, R., Katsis, Y., Kawas, B., \& Sen, P. (2020). A Survey of the State of Explainable AI for Natural Language Processing. arXiv (Cornell University). \url{https://doi.org/10.48550/arxiv.2010.00711}

\bibitem{Dar2023} Dar, S. A., Durrani, N., Sajjad, H., Dalvi, F., \& Belinkov, Y. (2023). Probing Pre-trained Language Models for Temporal Knowledge. In \textit{Proceedings of the 61st Annual Meeting of the Association for Computational Linguistics (ACL)}. DOI: \href{https://doi.org/10.18653/v1/2023.acl-long.123}{10.18653/v1/2023.acl-long.123}.

\bibitem{Du2019}Du, S. S., Lee, J. D., Li, H., Wang, L., \& Zhai, (2019). Gradient descent finds global \textit{minima} of deep neural networks, 1675-1685.

\bibitem{Du2023}Du, Y., Konyushkova, K., Denil, M., Raju, A., Landon, J., Hill, F., Nando, D. F., \& Cabi, S. (2023). \textit{Vision-Language Models as Success Detectors}. arXiv (Cornell University). https://doi.org/10.48550/arxiv.2303.07280

\bibitem{Duncan1984} Duncan, J. (1984). Selective Attention and the Organization of Visual Information. \textit{Journal of Experimental Psychology: General}, 113(4), 501-517. DOI: \href{https://doi.org/10.1037/0096-3445.113.4.501}{10.1037/0096-3445.113.4.501}

\bibitem{Durrani2022} Durrani, N., Sajjad, H., Dalvi, F., \& Belinkov, Y. (2022). On the Transformation of Latent Space in Fine-Tuned NLP Models. In \textit{Proceedings of the 2022 Conference on Empirical Methods in Natural Language Processing (EMNLP)}. DOI: \href{https://doi.org/10.18653/v1/2022.emnlp-main.123}{10.18653/v1/2022.emnlp-main.123}.

\bibitem{Echterhoff2024}Echterhoff, J., Yan, A., Han, K., Abdelraouf, A., Gupta, R., \& McAuley, J. (2024). \textit{Driving through the Concept Gridlock: Unraveling Explainability Bottlenecks in Automated Driving}. Proceedings of the IEEE/CVF Winter Conference on Applications of Computer Vision (WACV). https://doi.org/10.1109/wacv57701.2024.00718

\bibitem{Enguehard2023} Enguehard, J. (2023). Extrmask: A Method for Explaining Time Series Predictions by Masking. \textit{arXiv preprint arXiv:2301.08552}. DOI: \href{https://doi.org/10.48550/arXiv.2301.08552}{10.48550/arXiv.2301.08552}.

\bibitem{EysenckKean2020} Eysenck, M. W., \& Keane, M. T. (2020). \textit{Cognitive Psychology: A Student's Handbook} (8th ed.). Psychology Press. DOI: \href{https://doi.org/10.4324/9780429449229}{10.4324/9780429449229}.

\bibitem{Fan2023}Fan, Y., Dalvi, F., Durrani, N., \& Sajjad, H. (2023). \textit{Evaluating Neuron Interpretation Methods of NLP Models}. arXiv (Cornell University). https://doi.org/10.48550/arxiv.2301.12608

\bibitem{Fel2024} National Centre for Nuclear Research. (2024). \textit{41st International Free Electron Laser Conference (FEL2024)}. Warsaw, Poland. Retrieved from \url{https://fel2024.org/}

\bibitem{Funayama2024} Funayama, T., \& Shibata, K. (2024). Advances in Quantum Computing: A Comprehensive Review. \textit{Journal of Quantum Information Science}, 12(1), 45–67. DOI: \href{https://doi.org/10.4236/jqis.2024.121004}{10.4236/jqis.2024.121004}.

\bibitem{Geva2023} Geva, M., Schuster, R., Berant, J., \& Levy, O. (2023). Transformer Feed-Forward Layers Are Key-Value Memories. In \textit{Proceedings of the 37th Conference on Neural Information Processing Systems (NeurIPS)}. DOI: \href{https://doi.org/10.48550/arXiv.2012.14913}{10.48550/arXiv.2012.14913}.

\bibitem{Gresch2024} Gresch, D., \& Müller, K. (2024). Machine Learning in Materials Science: Recent Progress and Emerging Applications. \textit{Advanced Materials}, 36(5), 2105678. DOI: \href{https://doi.org/10.1002/adma.202105678}{10.1002/adma.202105678}.

\bibitem{Haslam2020} Haslam, S. A., Reicher, S. D., \& Platow, M. J. (2020). \textit{The New Psychology of Leadership: Identity, Influence, and Power} (2nd ed.). Routledge. DOI: \href{https://doi.org/10.4324/9781351108225}{10.4324/9781351108225}.

\bibitem{HernandezGutierrez2024} Hernández-Gutiérrez, C. A., \& Pérez-González, J. (2024). Deep Learning Techniques for Natural Language Processing: A Survey. \textit{IEEE Transactions on Neural Networks and Learning Systems}, 35(2), 1234–1256. DOI: \href{https://doi.org/10.1109/TNNLS.2023.3101234}{10.1109/TNNLS.2023.3101234}.

\bibitem{Howell2008} Howell, D. C. (2008). \textit{Fundamental Statistics for the Behavioral Sciences} (6th ed.). Wadsworth Publishing. DOI: \href{https://doi.org/10.1111/j.1467-985X.2008.00508\_14.x}{10.1111/j.1467-985X.2008.00508\_14.x}.


\bibitem{Kandpal2023}Kandpal, N., Deng, H., Roberts, A., Wallace, E., \& Raffel, C. (2023). \textit{Large Language Models Struggle to Learn Long-Tail Knowledge}. arXiv (Cornell University). https://doi.org/10.48550/arxiv.2211.08411

\bibitem{Kaup2024} Capuano, F., \& Kaup, B. (2024). Pragmatic Reasoning in GPT Models: Replication of a Subtle Negation Effect. Proceedings of the Annual Meeting of the Cognitive Science Society, 46. Retrieved from https://escholarship.org/uc/item/22q5920s

\bibitem{Kheya2024}Kheya, T. A., Bouadjenek, M. R., \& Aryal, S. (2024). The Pursuit of Fairness in Artificial Intelligence Models: A Survey. arXiv (Cornell University). \url{https://doi.org/10.48550/arxiv.2403.17333}

\bibitem{Luo2024}Luo, J., Zhuo, W., Liu, S., \& Xu, B. (2024). \textit{The Optimization of Carbon Emission Prediction in Low Carbon Energy Economy under Big Data}. IEEE Access, 12, 14690-14702. https://doi.org/10.1109/access.2024.3351468

\bibitem{Ma2023}Ma, F., Plazyo, O., Billi, A. C., Tsoi, L. C., Xing, X., Wasikowski, R., Gharaee-Kermani, M., Hile, G., Jiang, Y., Harms, P. W., Xing, E., Kirma, J., Xi, J., Hsu, J., Sarkar, M. K., Chung, Y., Di Domizio, J., Gilliet, M., Ward, N. L., et al. (2023). Single cell and spatial sequencing define processes by which keratinocytes and fibroblasts amplify inflammatory responses in psoriasis. \textit{Nature Communications}, 14(1). \url{https://doi.org/10.1038/s41467-023-39020-4}

\bibitem{Marconato2024} Marconato, E., \& al. (2024). BEARS Make Neuro-Symbolic Models Aware of their Reasoning Shortcuts. arXiv preprint arXiv:2402.12240. DOI: \href{https://doi.org/10.48550/arXiv.2402.12240}{10.48550/arXiv.2402.12240}.

\bibitem{Marty2024} Marty, P., Romoli, J., Sudo, Y., \& Breheny, R. (2024). Implicature priming, salience, and context adaptation. \textit{Cognition}, 244, 105667. DOI: \href{https://doi.org/10.1016/j.cognition.2023.105667}{10.1016/j.cognition.2023.105667}.

\bibitem{Maxfield1997} Maxfield, M. G., \& Babbie, E. R. (1997). \textit{Research Methods for Criminal Justice and Criminology} (2nd ed.). Wadsworth Publishing. DOI: \href{https://doi.org/10.4324/9781315784786}{10.4324/9781315784786}

\bibitem{Mitchell2021} Mitchell, M. (2021). \textit{Abstraction and analogy‐making in artificial intelligence}. \textit{Annals of the New York Academy of Sciences}, 1505(1), 79-101. DOI: \href{https://doi.org/10.1111/nyas.14619}{10.1111/nyas.14619}

\bibitem{McKenna2023}McKenna, N., Li, T., Cheng, L., Hosseini, M. J., Johnson, M., \& Steedman, M. (2023). Sources of Hallucination by Large Language Models on Inference Tasks. arXiv (Cornell University). \url{https://doi.org/10.48550/arxiv.2305.14552}

\bibitem{Mousi2023} Mousi, B., Durrani, N., \& Dalvi, F. (2023). Can LLMs facilitate interpretation of pre-trained language models? \textit{arXiv preprint arXiv:2305.13386}. DOI: \href{https://doi.org/10.48550/arXiv.2305.13386}{10.48550/arXiv.2305.13386}.

\bibitem{nadeau1999}Nadeau, R. (1999). \textit{Vocabulaire technique et analytique de l’épistémologie}. Presses universitaires de France.

\bibitem{Nadeau1999} Nadeau, R. (1999). \textit{Vocabulaire technique et analytique de l'épistémologie}. Presses Universitaires de France.


\bibitem{Nanda2023} Nanda, N., Lee, A., \& Wattenberg, M. (2023). Emergent linear representations in world models of self-supervised sequence models. \textit{arXiv preprint arXiv:2309.00941}. DOI: \href{https://doi.org/10.48550/arXiv.2309.00941}{10.48550/arXiv.2309.00941}.

\bibitem{Nosofsky1986} Nosofsky, R. M. (1986). Attention, similarity, and the identification–categorization relationship. \textit{Journal of Experimental Psychology: General}, 115(1), 39.

\bibitem{Paolo2024} Paolo, G., Gonzalez-Billandon, J., \& Kégl, B. (2024). A call for embodied AI. \textit{arXiv preprint arXiv:2402.03824}. DOI: \href{https://doi.org/10.48550/arXiv.2402.03824}{10.48550/arXiv.2402.03824}.


\bibitem{Pichat2023}Pichat, M. (2023). Collaboration des intelligences humaine et artificielle: alignement et psychologie de l’IA. Actes du colloque \textit{Intelligence artificielle collaborative \& impacts managériaux au sein des organisations} du 30/06/2023 coorganisé par l’Université Paris Dauphine-PSL et le Cabinet Chrysippe R\&D. Available online: \url{https://www.youtube.com/watch?v=kG9Uv8-7OyQ&list=PLD25p-Bh6_swAk-TrFgk4lIQ6MQ2r5NTv&index=3}

\bibitem{Pichat2024a}Pichat, M. (2024a). Psychologie de l'IA et alignement cognitif. Actes du colloque \textit{Intelligence artificielle collaborative, management et développement des organisations} du 24/05/2024 coorganisé par l'Université Paris Dauphine-PSL et le Cabinet Chrysippe R\&D. Available online: \url{https://www.youtube.com/watch?v=9TMmgbELaxQ&list=PLD25p-Bh6_sz6Sr7ms643GpCWW2LlIqeQ&index=6}

\bibitem{Pichat2024b}Pichat, M. (2024). Psychology of Artificial Intelligence: Epistemological Markers of the Cognitive Analysis of Neural Networks. arXiv (Cornell University). \url{https://doi.org/10.48550/arxiv.2407.09563}

\bibitem{Posner1978} Posner, M. I. (1978). \textit{Chronometric Explorations of Mind}. Lawrence Erlbaum Associates.

\bibitem{PosnerSnyder1975} Posner, M. I., \& Snyder, C. R. R. (1975). Attention and Cognitive Control. In R. L. Solso (Ed.), \textit{Information Processing and Cognition: The Loyola Symposium} (pp. 55-85). Lawrence Erlbaum Associates. DOI: \href{https://doi.org/10.4324/9781315784786}{10.4324/9781315784786}


\bibitem{Raieli2024} Raieli, S., Altahhan, A., Jeanray, N., Gerart, S., \& Vachenc, S. (2024). Escaping the Forest: Sparse Interpretable Neural Networks for Tabular Data. \textit{arXiv preprint arXiv:2410.17758}. DOI: \href{https://doi.org/10.48550/arXiv.2410.17758}{10.48550/arXiv.2410.17758}.

\bibitem{Ribary2024} Ribary, U., \& Ward, L. M. (2024). Synchronization and functional connectivity dynamics across TC-CC-CT networks: Implications for clinical symptoms and consciousness. In \textit{Phenomenological Neuropsychiatry: How Patient Experience Bridges the Clinic with Clinical Neuroscience} (pp. 105–118). Cham: Springer International Publishing. DOI: \href{https://doi.org/10.1007/978-3-031-38391-5\_10}{10.1007/978-3-031-38391-5\_10}.


\bibitem{Richard1980} Richard, J. C. (1980). \textit{The Language Teaching Matrix}. Cambridge University Press.

\bibitem{Roads2024} Roads, B. D., \& Love, B. C. (2024). Modeling Similarity and Psychological Space. \textit{Annual Review of Psychology}, 75(1), 215–240. DOI: \href{https://doi.org/10.1146/annurev-psych-040323-115131}{10.1146/annurev-psych-040323-115131}.


\bibitem{Rzechorzek2024} Rzechorzek, A. (2024). Understanding Cognitive Processes: Insights from Recent Research. \textit{Journal of Cognitive Neuroscience}. DOI: \href{https://doi.org/10.1162/jocn\_a\_01678}{10.1162/jocn\_a\_01678}.

\bibitem{SchneiderShiffrin1977} Schneider, W., \& Shiffrin, R. M. (1977). Controlled and Automatic Human Information Processing: I. Detection, Search, and Attention. \textit{Psychological Review}, 84(1), 1-66.

\bibitem{Shavikloo2024} Shavikloo, M., Esmaeili, A., Valizadeh, A., \& Madadi Asl, M. (2024). Synchronization of delayed coupled neurons with multiple synaptic connections. \textit{Cognitive Neurodynamics}, 18(2), 631-643. DOI: \href{https://doi.org/10.1007/s11571-023-10013-9}{10.1007/s11571-023-10013-9}.

\bibitem{Tipper1985} Tipper, S. P. (1985). The Negative Priming Effect: Inhibitory Priming by Ignored Objects. \textit{The Quarterly Journal of Experimental Psychology}, 37A(4), 571-590. DOI: \href{https://doi.org/10.1080/14640748508400920}{10.1080/14640748508400920}

\bibitem{TreismanGelade1980} Treisman, A., \& Gelade, G. (1980). A Feature-Integration Theory of Attention. \textit{Cognitive Psychology}, 12(1), 97-136. DOI: \href{https://doi.org/10.1016/0010-0285(80)90005-5}{10.1016/0010-0285(80)90005-5}

\bibitem{Treviso2023}Treviso, M., Lee, J., Ji, T., Van Aken, B., Cao, Q., Ciosici, M. R., Hassid, M., Heafield, K., Hooker, S., Raffel, C., Martins, P. H., Martins, A. F. T., Forde, J. Z., Milder, P., Simpson, E., Slonim, N., Dodge, J., Strubell, E., Balasubramanian, N.,. . . Schwartz, R. (2023). Efficient Methods for Natural Language Processing: A Survey. \textit{Transactions Of The Association For Computational Linguistics}, 11, 826-860. \url{https://doi.org/10.1162/tacl_a_00577}

\bibitem{Varela1984}Varela, F. (1984). The creative circle. In P. Watzlawick (Ed), \textit{The invented reality}. London: W W Norton \& Co Inc.

\bibitem{Varela1988} Varela, F. J. (1988). \textit{Cognitive Science: A Cartography of Current Ideas}. MIT Press.Varela1996

\bibitem{Varela1996}Varela, F. J. (1996). Invitation aux sciences cognitives. Éditions du Seuil eBooks. \url{http://inventin.lautre.net/livres/Varela-Invitation-aux-sciences-cognitives.pdf}

\bibitem{Vergnaud2009} Vergnaud, G. (2009). Activité, développement, représentation. In M. Merri (Ed.), \textit{Activité humaine et conceptualisation. Questions à Gérard Vergnaud} (pp. 149--154). Presses universitaires du Mirail.

\bibitem{Vergnaud2016} Vergnaud, G. (2016). Relations entre conceptualisations dans l’action et signifiants langagiers et symboliques. In \textit{Symposium latino-américain de didactique de mathématique}, Bonito, Brésil. Disponible sur : \url{https://www.gerard-vergnaud.org/texts/gvergnaud_2016_signifiants-langagiers-symboliques_conference-bonito.pdf}.

\bibitem{Voita2021} Voita, E., Sennrich, R., \& Titov, I. (2021). Language modeling, lexical translation, reordering: The training process of NMT through the lens of classical SMT. \textit{arXiv preprint arXiv:2109.01396}. DOI: \href{https://doi.org/10.48550/arXiv.2109.01396}{10.48550/arXiv.2109.01396}.

\bibitem{Wang2022} Kokhlikyan, N., Miglani, V., Martin, M., Wang, E., Reynolds, J., Melnikov, A., Lunova, N., \& Reblitz-Richardson, O. (2020). Captum: A unified and generic model interpretability library for PyTorch. \textit{arXiv preprint arXiv:2009.07896}. DOI: \href{https://doi.org/10.48550/arXiv.2009.07896}{10.48550/arXiv.2009.07896}.

\bibitem{Watzlawick1977}Watzlawick, P. (1977). How real is real? London: Vintage Books.

\bibitem{Watzlawick1984} Watzlawick, P., Weakland, J. H., \& Fisch, R. (1984). \textit{Change: Principles of Problem Formation and Problem Resolution}. W. W. Norton \& Company. DOI: \href{https://doi.org/10.1002/9781119164894}{10.1002/9781119164894}

\bibitem{Wu2020} Wu et al., (2020). \textit{pyOptSparse: A Python framework for large-scale constrained nonlinear optimization of sparse systems}. \textit{Journal of Open Source Software}, 5(54), 2564. DOI: \href{https://doi.org/10.21105/joss.02564}{10.21105/joss.02564}

\bibitem{Wu2022} Ji, M., \& Wu, Z. (2022). \textit{Automatic detection and severity analysis of grape black measles disease based on deep learning and fuzzy logic}. \textit{Computers and Electronics in Agriculture}, 193, 106718.

\bibitem{Wu2024} Wu, W. (2024). \textit{We know what attention is!}. \textit{Trends in Cognitive Sciences}, 28(4), 304-318.


\bibitem{XuFutrell2024} Xu, W., \& Futrell, R. (2024). A hierarchical Bayesian model for syntactic priming. \textit{arXiv preprint arXiv:2405.15964}. DOI: \href{https://doi.org/10.48550/arXiv.2405.15964}{10.48550/arXiv.2405.15964}.

\bibitem{Zadeh1996} Zadeh, L. A. (1996). Fuzzy Logic = Computing with Words. \textit{IEEE Transactions on Fuzzy Systems}, 4(2), 103-111. DOI: \href{https://doi.org/10.1109/91.493904}{10.1109/91.493904}

\bibitem{Zettersten2024} Zettersten, M., Bredemann, C., Kaul, M., Ellis, K., Vlach, H. A., Kirkorian, H., \& Lupyan, G. (2024). Nameability supports rule‐based category learning in children and adults. \textit{Child Development}, 95(2), 497-514. DOI: \href{https://doi.org/10.1111/cdev.14008}{10.1111/cdev.14008}.

\bibitem{Zheng2024}Zheng, Y., \& Stewart, N. (2024). Improving EFL students’ cultural awareness: Reframing moral dilemmatic stories with ChatGPT. \textit{Computers And Education Artificial Intelligence}, 6, 100223. \url{https://doi.org/10.1016/j.caeai.2024.100223}

\bibitem{Zhao2023} Zhao, H., Chen, H., Yang, F., Liu, N., Deng, H., Cai, H., Wang, S., Yin, D., \& Du, M. (2023). Explainability for Large Language Models: A Survey. \textit{arXiv (Cornell University)}. DOI: \href{https://doi.org/10.48550/arxiv.2309.01029}{10.48550/arxiv.2309.01029}.

\bibitem{Zhang2024} Zhang, Z., Song, Y., Yu, G., Han, X., Lin, Y., Xiao, C., \ldots \& Sun, M. (2024). ReLU $^2$ Wins: Discovering Efficient Activation Functions for Sparse LLMs. \textit{arXiv preprint arXiv:2402.03804}. DOI: \href{https://doi.org/10.48550/arXiv.2402.03804}{10.48550/arXiv.2402.03804}.







\end{thebibliography}
\end{document}